\documentclass[sigconf]{acmart}

\AtBeginDocument{%
  \copyrightyear{2022 CC-BY 4.0 Copyright held by the owner/author(s). Publication rights licensed to} 
\acmConference[FAccT '22]{2022 ACM Conference on Fairness, Accountability, and Transparency}{June 21--24, 2022}{Seoul, Republic of Korea}
\acmBooktitle{2022 ACM Conference on Fairness, Accountability, and Transparency (FAccT '22), June 21--24, 2022, Seoul, Republic of Korea}
\acmPrice{15.00}
\acmDOI{10.1145/3531146.3533156}
\acmISBN{978-1-4503-9352-2/22/06}

  \providecommand\BibTeX{{%
    \normalfont B\kern-0.5em{\scshape i\kern-0.25em b}\kern-0.8em\TeX}}}

\usepackage[T1]{fontenc}
\usepackage[utf8]{inputenc}
\usepackage{url}
\usepackage{breakurl}
\begin{document}
%%
%% The "title" command has an optional parameter,
%% allowing the author to define a "short title" to be used in page headers.
\title[German AI Start-Ups and "AI Ethics"]{German AI Start-Ups and “AI Ethics”: Using A Social Practice Lens for Assessing and Implementing Socio-Technical Innovation}

%%
%% The "author" command and its associated commands are used to define
%% the authors and their affiliations.
%% Of note is the shared affiliation of the first two authors, and the
%% "authornote" and "authornotemark" commands
%% used to denote shared contribution to the research.
\author{Mona Sloane}
 \email{mona.sloane@uni-tuebingen.de}
 \email{mona.sloane@nyu.edu}
 \orcid{1234-5678-9012}
\affiliation{%
  \institution{University of Tübingen}
    \city{Tübingen}
     \postcode {72074}
     \country{Germany}}
     \affiliation{%
  \institution{New York University}
  \city{New York City}
    \country{USA}
}
\author{Janina Zakrzewski}
\email{janina.zakrzewski@uni-tuebingen.de}
\affiliation{%
  \institution{University of Tübingen}
  \city{Tübingen}
    \postcode{72074}
    \country{Germany}
}

%%
%% By default, the full list of authors will be used in the page
%% headers. Often, this list is too long, and will overlap
%% other information printed in the page headers. This command allows
%% the author to define a more concise list
%% of authors' names for this purpose.
\renewcommand{\shortauthors}{Sloane and Zakrzewski}

%%
%% The abstract is a short summary of the work to be presented in the
%% article.
\begin{abstract}
  The current AI ethics discourse focuses on developing computational interpretations of ethical concerns, normative frameworks, and concepts for socio-technical innovation. There is less emphasis on understanding how AI practitioners themselves understand ethics and socially organize to operationalize ethical concerns. This is particularly true for AI start-ups, despite their significance as a conduit for the cultural production of innovation and progress, especially in the US and European context. 

This gap in empirical research intensifies the risk of a disconnect between scholarly research, innovation and application. This risk materializes acutely as mounting pressures to identify and mitigate the potential harms of AI systems have created an urgent need to rapidly assess and implement socio-technical innovation focused on fairness, accountability, and transparency.

In this paper, we address this need. Building on social practice theory, we propose a framework that allows AI researchers, practitioners, and regulators to systematically analyze existing cultural understandings, histories, and social practices of “ethical AI” to define appropriate strategies for effectively implementing socio-technical innovations. We argue that this approach is needed because socio-technical innovation “sticks” better if it sustains the cultural meaning of socially shared (ethical) AI practices, rather than breaking them. By doing so, it creates pathways for technical and socio-technical innovations to be integrated into already existing routines.

Against that backdrop, our contributions are threefold: (1) we introduce a practice-based approach for understanding “ethical AI”; (2) we present empirical findings from our study on the operationalization of “ethics” in German AI start-ups to underline that AI ethics and social practices must be understood in their specific cultural and historical contexts; and (3) based on our empirical findings, suggest that “ethical AI” practices can be broken down into principles, needs, narratives, materializations, and cultural genealogies to form a useful backdrop for considering socio-technical innovations. We conclude with critical reflections and practical implications of our work, as well as recommendations for future research.

\end{abstract}

%%
%% The code below is generated by the tool at http://dl.acm.org/ccs.cfm.
%% Please copy and paste the code instead of the example below.
%%



%%
%% Keywords. The author(s) should pick words that accurately describe
%% the work being presented. Separate the keywords with commas.

%%
%% This command processes the author and affiliation and title
%% information and builds the first part of the formatted document.

\maketitle
\section{Introduction}
Addressing algorithmic harms and implementing mechanisms for enhancing fairness, accountability and transparency in AI systems is a pressing challenge for researchers, practitioners, and policymakers alike. New regulatory frameworks on transnational, national, federal, and local levels create an urgent need to rapidly assess and implement technical and socio-technical innovation that fosters fairness, accountability, and transparency. This is particularly true for AI start-ups, which are often treated as conduits for the cultural production of innovation and the acceleration of economic progress, especially in the US and European contexts \cite{sloane_ethics_2022}. They face the same regulatory pressures as large organizations but cannot fall back onto comparable resources and knowledge for capacity building in what practitioners often call “AI ethics” \cite{greene_better_2019}. 

To date, we know little about how AI start-ups put into practice or operationalize “AI ethics”. This gap is indicative of a larger split in the field of AI ethics: There is a rapidly growing body of work that focuses on technical innovation via computational interpretations of ethical concerns, normative frameworks, and socio-technical innovation which is outpacing the production of empirical research on the professional practices of AI development. The most recent technical innovations in the AI ethics space include, for example, tactics for fair clustering \cite{abbasi_fair_2021} and producing intersectionally fair rankings \cite{yang_causal_2021}, as well as testing the probabilistic fairness of pre-trained logistic classifiers \cite{taskesen_statistical_2021}, assessing “leave-one-out unfairness” \cite{black_leave-one-out_2021}, or measuring robustness bias \cite{nanda_fairness_2021}, among many others. 

In the academe, normative approaches that have been introduced include the introduction of technomoral virtues \cite{vallor_technology_2016}, contextual integrity \cite{nissenbaum_privacy_2009}, or artificial morality \cite{allen_artificial_2005}. In practice, normative concepts such as utilitarianism continue to play a significant role in AI applications, especially self-driving vehicles \cite{jafarinaimi_our_2018, karnouskos_role_2021}. Similarly relevant are normative AI ethics codes that seek to influence behavior of AI professionals \cite{ryan_artificial_2020}. 

Socio-technical innovations to help foster fairness, accountability, and transparency include model cards for model reporting \cite{mitchell_model_2019} and datasheets for datasets \cite{gebru_datasheets_2021}, responsible data management \cite{stoyanovich_responsible_2020}, counterfactuals \cite{pearl_causality_2009, morgan_counterfactuals_2015, kusner_counterfactual_2017, kilbertus_avoiding_2017, karimi_algorithmic_2021, barocas_hidden_2020}, using measurement theory to render assumptions in mathematical models visible \cite{jacobs_measurement_2021, milli_optimizing_2021}, deploying structural causal models \cite{kacianka_designing_2021} or introducing traceability \cite{kroll_outlining_2021} to define, assess and improve models of accountability, nutritional labels \cite{holland_dataset_2018, stoyanovich_nutritional_2019}, algorithmic impact assessment \cite{metcalf_algorithmic_2021, reisman_algorithmic_2018, kaminski_algorithmic_2019, selbst_disparate_2017}, or algorithmic auditing \cite{kasy_fairness_2021, raji_closing_2020, koshiyama_towards_2021, bandy_problematic_2021, brown_algorithm_2021, sloane_silicon_2021, rhea_external_nodate}. 

While this important body of work is fast expanding, less is known about how professionals treat and enact AI ethics vis-à-vis these innovations. Even though the fairness, accountability and transparency (FAccT) community increasingly turns towards an integrative understanding of \textit{socio}-technical systems, the emphasis on technology lingers and supplants advancements in understanding how “the social” in “socio-technical” systems actually pans out, especially in the context of the cultural production of AI innovation at large. The epistemological neglect of social practice and the cultural context of AI can perpetuate a mentality of framing the social impact of AI as a purely technical problem\cite{vakkuri_this_2020}. This technical framing can foster a disconnect to cultural context and social practice and lead to mitigation strategies only hovering at the surface. Prominent examples of that are different forms of “-washing” such as ethics-washing \cite{greene_better_2019, ochigame_invention_2019, sloane_inequality_2019, wagner_ethics_2018}, participation-washing \cite{sloane_participation_2020}, potential diversity, equality, and inclusion (DEI)-washing \cite{howard_council_2020}, or woke-washing \cite{dowell_woke-washing_2020}.
  
The point of departure for this paper is that this epistemological gap is the \textit{actual} crux of what is often described as the polarization between techno-solutionism and techno-criticism: failure to systematically understand the cultural contexts and social practices AI gets folded into across domains intensifies the risk of a disconnect between scholarly research and application at a time when convergence is much needed to effectively and sustainably address AI harms and comply with new regulatory regimes. This issue surfaces concretely in the fairness, accountability, and transparency space of an issue of adoption \cite{kroll_outlining_2021}, or integration.\footnote{Of course, there also is a bigger political issue that comes with prioritizing quantitative ways of knowing and describing the world over qualitative approaches. For example, such an asymmetry systematically prevents genuine intersectional approaches  \cite{crenshaw_mapping_1991, hammack_jr_intersectional_2018}, which center lived experience, to enter the frame, and therefore can be seen as perpetuating oppression. Those considerations, however, are out of scope for this paper.} What seems to be missing is a framework that can meaningfully connect the rich insights from qualitative works and conceptual advancements that have focused on studying the intersection of algorithmic systems and social life, particularly with regards to the professions.\footnote{This body of work is expansive and continues to grow. Important works include research on the gig economy \cite{gray_ghost_2019, rosenblat_rosenblat_2018, cameron_making_2021, ticona_trusted_2018}, journalism \cite{christin_metrics_2020, petre_all_2021}, content moderation \cite{roberts_behind_2019}, or hiring \cite{dencik_regimes_2021}.}
 
In this paper, we address this need. We present empirical findings from our study on the cultural interpretation, history, and operationalization of “ethics” in German AI start-ups. We use the analysis of this material to argue that AI ethics must be understood in their specific cultural, social, and historical contexts. We base this work on a practice-based approach for understanding “ethical AI” to then present an “anatomy of AI ethics” in German AI start-ups that breaks down “ethical AI” practices into principles, needs, narratives, materializations, and cultural genealogies, all of which we illustrate with data from the field. 
 
To discuss applicability of this framework beyond the academe, we then translate our conceptual work into a research and innovation assessment guide that allows AI researchers, practitioners, and regulators to systematically analyze existing cultural understandings, histories, and social practices of “ethical AI” to define appropriate strategies for effectively implementing socio-technical innovations that emerge in the FAccT field. This guide is comprised of two sets of questions that can help researchers and practitioners conduct their own qualitative work mapping the existing cultural understandings, histories, and social practices and use this analysis as a method of assessing technical and socio-technical innovations vis-à-vis their potential for adoption. We close this paper with critical reflections on our own work and recommendations for future research.

\subsection{Why is understanding AI ethics as social practice necessary?}

Working to understand the practical accomplishment of “ethical work” in technology organizations is not a new endeavor. Fruitful, and empirically grounded, approaches have been proposed, notably Ziewitz’s work on SEO consultants and the “ethicality of optimization” \cite{ziewitz_rethinking_2019,rakova_where_2021} work on the impact of organizational culture and structure on the effectiveness of responsible AI initiatives in practice, or Madaio et al.’s research \cite{madaio_assessing_2022} on organizational factors shaping the work on AI fairness. 

Here, however, we are taking the cue from Metcalf and Moss' \cite{metcalf_owning_2019} notion of “ethics owners”, who define ethics owners in their report \cite{moss_ethics_2020} as "Silicon Valley professionals tasked with managing the ethical footprints of tech companies”, and who handle “challenging ethical dilemmas with the tools of tech management and within familiar tech company structures”. But rather than focusing on individual ethics roles and responsibilities, we want to center the question of how meaning is ascribed to “ethics” in relation to the wider professional practice of AI design, and how this meaning stabilizes. Within that, we are looking to make space for the cultural specificity of how this dynamic unfolds, as well as its material dimension --- we are less concerned with individualized social processes. Therefore, we engage with the analytically potent notion of social practice theory \cite{bourdieu_outline_1977, shove_design_2007,hui_matters_2017, hui_variations_2017, reckwitz_toward_2002, reckwitz_kreativitat_2016, schatzki_social_1996, schatzki_site_2002, schatzki_practice_2001}, a framework that decenters the individual and focuses on a practice as a unit of inquiry  \cite{heidenstrom_utility_2021}. 

Specifically, we build on Shove, Pantzar, \& Watson’s \cite{shove_dynamics_2012} framework that focuses on the \textit{dynamics} of social practice as a way of charting and understanding patterns of stability and change. They suggest that social practices are comprised of the three elements: \textit{meanings}, \textit{competences}, and \textit{materials}. \textit{Meanings} designate the social and symbolic significance of participation in any given moment, including emotional knowledge; \textit{competences} are skills, multiple forms of shared understanding and the practical knowledgeability; \textit{materials} are objects, infrastructures, tools, hardware and so on, as well as the body itself \cite{shove_dynamics_2012}. These elements are individually distributed and combined but inform each other. A social practice stabilizes when the three elements are linked up, whereby this stabilization is not static, because elements change as linkages between them are continually made and remade and practices stabilize (or disintegrate) across time and space. The making, breaking, and re-making of links between the elements is “transformative” in that it (re-)shapes those elements that do not disintegrate. Important, here, is the idea that continuities shape competences, materials, and meanings, and how they link up \cite{shove_dynamics_2012}. Shove, Pantzar, \& Watson \cite{shove_dynamics_2012} use the case of the social practice of (car-)driving to work this out, discussing how the material design of a car is a continuation of the design of a horse carriage, rather than a radical new design, and that driving on the left-hand side of a road (in the UK) relates to swords typically being held on the right hand.

What is important to note in this context is that Shove, Pantzar, \& Watson \cite{shove_dynamics_2012} describe the connection of various social practices in terms of “bundles” and “complexes” of practice, as well as circuits of reproduction, that have “emergent lives of their own” \cite{shove_dynamics_2012}. Whilst we do not contend that, we are interested in what that means in the context of the body of a profession, and the profession of AI design and the field of German AI start-ups specifically. In other words, we are concerned with an interpretation and a use of social practice theory that has a more bounded focus, that is more closely tied to the research focus on organizations, and reflects the insight that was gathered from the empirical data. 

Therefore, we use social practice theory in conjunction with our empirical data to develop an “anatomy” of AI ethics. The connecting thread between the two is a concern for stabilization: the anatomy of AI ethics is comprised of a “skeleton” and of “soft tissue” which together stabilize AI ethics in German start-ups. The “skeleton” of AI ethics is comprised of five connected and mutually constitutive AI ethics elements that broadly map onto the characteristics of the social practice elements of competencies, meanings, and materials, but are more nuanced (see discussion below): principles, needs, narratives, ethics materials, and cultural genealogy. Whilst these five elements scaffold AI ethics, they are brought to life – in a specific context – by the “soft tissue” of this anatomy: the way in which principles, needs, narratives, ethics materials, and cultural genealogy are enacted in an interconnected way. In other words, the “skeleton” is the What of (the stabilization of) AI ethics in German start-ups, and the “soft tissue” is the How. Below, we illustrate this dynamic by discussing the enactment of the principles “Mitbestimmung” (co-determination) and “Verantwortung” (responsibility).\footnote{To date, our analysis shows 10 principles overall: “Mitbestimmung” (co-determination), “Verantwortung” (responsibility), “Ehrlichkeit” (honesty), “Aufklärung” (education / enlightenment), “Austausch” (exchange), “Fortbildung” (training), “Absicherung” (coverage), “Ordnung” (order), “Forschung” (research), and “Miteinander” (togetherness). Discussing all of these in this paper, however, is out of scope. We therefore decided to only showcase the two most prominent principles to show how the elements are connected.}  It is important to note that the anatomy frame is not a replacement for the social practice theory approach, it is an expansion thereof. 

As demonstrated in the discussion part of this paper, we argue that using social practice theory in this way is particularly fruitful for researchers and practitioners who wish to consider where and how technical and socio-technical innovation on fairness, accountability, and transparency can be best “linked into” existing social practices of AI design and stabilized as an integral element of them - or in other words: so that they “stick” better.

\section{Data Collection, Analysis and Limitations}
This paper reports on empirical findings that are part of a larger qualitative study on the operationalization of ethics in German AI start-ups.\footnote{The project is housed at the Tübingen AI Center, University of Tübingen, Germany, and the Principal Investigator is Mona Sloane.} The study aims to map out and understand the plurality of social assumptions and historical continuities that give shape to how German AI start-ups put “ethics” into practice as an organizational responsibility and as a cultural understanding. Guiding research questions are: What is the plurality of meanings of “ethics” among German AI start-ups? What are the basic social assumptions and cultural processes that underpin “ethics” in German AI start-ups? What are the principles, processes, and practices that materialize “ethics” in German AI start-ups?

\subsection{Data Collection and Analysis}

Because this study is focused on researching articulations and enactments of “ethics” through the lens of AI practitioners (in this case in German AI start-ups), a qualitative research approach in the form of semi-structured interviews was chosen.\footnote{The original research proposal also encompassed participant observation in selected AI start-ups and accelerators, primarily in locations where there are AI start-up hubs, such as the Cybervalley in Tübingen, and various AI campuses in Berlin. Due to the Covid-19 pandemic, however, in-person research was suspended and the research team had to focus on conducting interviews virtually. It is important to note that due to the lack of in-person engagement, it was extremely difficult to source research participants and grow the data pool.} Research participants were sourced through the German Federal Association of AI Businesses (KI Bundesverband), the University of Tübingen, various AI industry event pages, and personal networks of the research team. All companies that self-declared to be a start-up, and that offered AI technology, were included in the sample for outreach. As of December 2021, the research team had conducted 64 interviews. 64\% of the individuals that were interviewed identified as AI start-up founders or co-founders, 30\% as AI start-up employees. The remaining 6\% were representatives of AI start-up associations or accelerators. 

For each interview, an interview protocol encompassing seven topics formed the basis for the conversation. These themes were derived from a scoping literature review on AI ethics (see section 1), and the empirical focus developed for the study. The themes were: start-up background, which included questions around the AI that the company built or used; ethics within the start-up, which included questions about how the interviewee defines ethics, as well as how ethics - as per their own definition - plays a role in the organization; ethics pertaining to technology design, including questions of how ethics are measured, if at all; ethics and AI generally, including questions about the social role of AI, and AI innovation in the context of inequality; the German start-up landscape, including questions around availability of government funding vs. venture capital, geographic or thematic AI innovation clusters, and the general understanding of “ethics” within the German AI industry; professional ethics, including the question of professional responsibility and accountability; and regulation, including questions about data protection, German and European approaches to AI regulation, and their relationship to ethics. 

As per IRB protocol,\footnote{This study was approved by the IRB of the University of Tübingen (Ethikkommission).} all research participants had to sign a consent form prior to the interview.  Of the 64 interviews 60 were conducted in German, 4 were conducted in English. All interviews were recorded and then transcribed using transcription software. Transcripts were cleaned manually and anonymized and entered into the transcription software Atlas.ti, where all interview material was coded.\footnote{All data is stored securely on password-protected servers. Codebooks are stored separately from interview data. As per German data protection regulation, the code list will be destroyed by 2023. The anonymized data will be kept for at least 10 years.} 

The data analysis followed a grounded theory approach. Grounded theory \cite{clarke_situational_2005, glaser_discovery_2017, charmaz_constructing_2006} is an approach that generates theory from data in an iterative way. This approach was chosen as it is most appropriate for studies that are based on open-ended research questions. It was also chosen because there is little known about the phenomenon of how actors within German AI start-ups specifically construct and enact “ethics”. Grounded theory was also chosen because it allowed for flexibility in the research process. Specifically, the grounded theory approach allowed for iteration in both data collection and analysis. In terms of data collection, it allowed shifting emphasis in interviews while sticking to the interview protocol. Initial coding followed the themes in the interview protocol, as well as the social practice theory lens deployed in this study (with a particular focus on coding for the elements). As coding progressed, clearer and more nuanced patterns emerged, as well as the relationship between them. We learned, for example, that the element of “meaning” bifurcates into needs, narratives, and cultural genealogy, all of which are culturally and historically specific. As data collection progressed and researchers gained more knowledge and understanding of the field, questions could be asked in a more targeted way. For example, the question of how “ethics” are materialized within the company initially was asked in an abstract way and later became a series of questions asking the research participant to mentally visualize the company, its environment and relevant actors, projects, processes, and how they are connected, and then to articulate where they “see” ethics in that visual. In terms of analysis, it allowed letting the data point to analytical foci and theorizations that were not expected when the study was originally designed. The grounded theory developed here – the anatomy of AI ethics – emerged at this juncture, allowing us to move from Shove, Pantzar, \& Watson’s notion of clustered and bundled practices, to the profession- and organization-focused concept of the anatomy of AI ethics. 

\subsection{Limitations}

There are a number of limitations of this study. First, this study depends on voluntary participation of participants and resulting self-selection of participants and a data set that is likely to be biased. In this case towards start-ups that were part of the ecosystems that the research team was able to tap into, towards start-ups that are already thinking about ethics and/or want to participate in the AI ethics discourse, or start-ups that have the “ethical” stance to support research (see below). Due to self-selection, the data set is also biased towards founders and co-founders rather than employees or freelancers. Second, restrictions on in-person research due to the COVID-19 pandemic severely limited the scope of the originally envisioned research. Specifically, the research team were not able to conduct in-person research via participant observation or more informal interviews (“water cooler conversations”) on site – methods that are essential for observing and recording social practices \cite{schmidt_sociology_2017} – and had to rely on the cumbersome process of formally scheduling and conducting interviews on video platforms. Because in interviews, especially virtually conducted interviews, practices are not directly observed but captured through the reporting of those enacting them, some scholars have questioned their viability for social practice research \cite{schmidt_sociology_2017, bourdieu_outline_1977}. Others, however, have rejected participant observation as the “golden standard” of social practice research and have argued for the benefit of interviews as producing equally relevant knowledge on practices, not least because narratives can then enter the analytical frame, and because interviews themselves are a social practice \cite{halkier_questioning_2017, nicolini_practice_2017}.\footnote{It must be noted that the overwhelming majority of interviewees reported that their entire organization was working remotely at the time of the interview. Some start-ups did not even have office space but were fully remote. Many corporate and non-corporate organizations have fully embraced a remote or hybrid office structure. Against that backdrop, social practice researchers will have to fundamentally rethink what it means to collect data on social practice.} For this research, we have taken the cue from the latter group, and argue that interviews are an appropriate method for conducting research on social practices, falling in line with scholars who have called for a “method pluralism” in social practice research \cite{littig_combining_2017}. Third, this study is limited, much like any other empirical research, through the powerful position the researchers take on as gatekeepers of both data collection and data analysis, requiring reflexivity of the researchers’ own positions and perspectives. In this case, this means acknowledging that both researchers identify as middle-class, European, white, native Germans with proficiency in English,\footnote{It must be noted that proficiency in both German and English enabled the research team to delve into relevant scholarly literature in both languages.} and with degrees from elite institutions (among other identifiers), characteristics they share with many of the research participants. While these shared identifiers may have facilitated access to some research participants, predominantly founders and co-founders, it may have prevented access to others, predominantly employees.

\section{The Anatomy of “AI Ethics” in German AI Start-Ups}

\subsection{The Skeleton}

\subsubsection{Principle/s} 
The social practices that constitute “ethics” in German AI start-ups are anchored by distinct principles that permeate across the field of German AI start-ups at large and that serve as actionable frameworks and actions within the organization. These principles have a central function in that the other elements - needs, narratives, ethics materials and cultural genealogy - are articulated \textit{in reference} to the principles. The principles presented themselves not as what is commonly understood as human values or as virtues per se, or as what has been critiqued as norms and methods that have little bearing on actual practice  \cite{mittelstadt_principles_2019}, but as frameworks that directly affected collective behavior, particularly within an organization. As such, they are closely related to what has been described as “business ethics”, which is “focused on the human values to coordinate activity inside a business operation” \cite{moss_ethics_2020}. In that sense, principles serve as framing devices for what is considered “ethical” practice within an organization and, perhaps more importantly what is not, and how this practice is organized, and accounted for  \cite{korte_einfuhrung_2016}.  

The principles we identified were both aspirational and concrete in that they represented ethical ideals a company had set for itself, for example creating and maintaining a culture of responsibility, as well as concrete actions and activities that materialized that ideal (see “ethics materials” below). They were underpinned and carried through by distinct narratives, they were serving particular needs of the organization and the AI community of practice at large, they had clearly identifiable cultural and historic roots, and they materialized as various socio-technical arrangements, roles, responsibilities, and so on. 

The principles that clearly emerged in the data showed clearly identifiable roots in German history and culture. We therefore chose to retain the original German terminology to describe them.\footnote{ It is well known that language and knowledge shape each other \cite{coupland_relation_1997, kay_what_1984}, and that certain terms cannot be readily translated, partially because they provide a certain vagueness that serves as an umbrella term for multiple interpretations which are culturally specific. Whilst ontologically difficult to pin down, these terms can play a central role for ordering social life. For example, Bille \cite{bille_hazy_2015, bille_vagueness_2019} argues that the Danish (and Norwegian) word “hygge” does not just mean “cosy”, but describes an atmosphere, a whole way of being in the world, and of being with other people. “Hygge” is aspirational, something to strive for, as well as real and distinctly materializes in social situations and is immediately recognizable by those familiar with this concept. It often stands in for what is “typical” Nordic culture \cite{linnet_money_2011}. The Dutch notion of “gezellig” is comparable, yet different, connoting cosyness and comfort, but with a distinct emphasis on conviviality over solitude. The principles that emerged from how our research participants described the role of “ethics” in their organizations showed a similar simultaneousness of vagueness and specificity that appeared to be best graspable by retaining its original terminology.} This is as much a methodological strategy as it is an analytical stance, because it centers the idea that what counts as “AI ethics” is not only culturally specific, but stabilizes as part of the wider social practice of AI design within a field~\cite{sloane_use_nodate}. It is important to note that principles co-emerge and overlap, rather than being strictly separate from one another, and that they can encompass a wide range of concerns and activities. The analysis below should, therefore, be read as the beginning of an in-depth analysis and mapping of AI ethics concerns in German start-ups which will grow as the data analysis progresses.

\subsubsection{Needs}
Our data showed that “ethics” in German AI start-ups materialized not altruistically, but served to meet a wider range of (business) needs that were internal to the AI start-up organization. These needs arose from very specific social, political, and commercial contexts, such as German data protection regulation. In the “soft tissue” section, we describe how these contexts created pressures, issues, and concerns that materialized concretely as problems in and for the organizations we studied, such as issues around talent scarcity, transparency mandates, or client acquisition and retention. These problems were addressed through \textit{the way in which} the aforementioned principles were enacted within the organization. For example, the need for scarce talent became an ethics problem as AI start-up founders found technical talent fastidious with workplace choices and preferring organizations that allowed for their individual ethical concerns to be heard and integrated into decision-making processes, such as around client selection.

\subsubsection{Narratives}
While principles depict broader AI ethics frameworks that work to influence collective behavior, and needs represent concrete problems that organizations “solved” through ethics, narratives were concrete stories that informed the organizational and social logics that underpinned the principles and their enactment. Narratives are “ensemble [...] of texts, images, spectacles, events and cultural artefacts that ‘tell a story’” \cite{bal_narratology_2017}. Narratives are both more concrete and elaborate than principles and they play an important role for how individuals, communities, and organizations make sense of and act upon the social worlds they are involved in~\cite{gubrium_new_1997}. Typically, the role of narratives is to justify the \textit{how} principles are enacted, and needs are met. Recent research has shown that narratives around AI in particular play an important role in shaping public opinion on technology, as well as technology innovation and regulation \cite{cave_portrayals_2018, cave_ai_2020, singler_existential_2019}. Importantly, they often serve to rationalize social and racial hierarchies and the status quo  \cite{benjamin_race_2019, browne_dark_2015, weheliye_habeas_2014}.

The narratives that we captured had a distinct function for the organizations we researched.\footnote{The narratives in the “soft tissue” section are not taken from the dataset verbatim, but are aggregates of statements and stories.} Sometimes, they presented in the form of if-then statements that rationalized particular stabilizations of the ethics principles. For example, a narrative that sustained how “Verantwortung” (responsibility) was enactacted was that individuals had a responsibility to articulate their personal “ethical” concerns, for example about projects or clients, and that this dynamic would create a generally more ethical organization (“\textit{If} everybody takes seriously their responsibility to articulate their personal ethical concerns, then the company will be more ethical.”). It is important to note that there were multiple and continually emerging narratives connected to each of the principles we identified, but they all connected directly to the needs of the organization.

\subsubsection{ Ethics Materials} 
Principles, needs, narratives, and cultural genealogies concretely came to bear through what we call “ethics materials”: concrete objects, processes, roles, tools or infrastructures focused on “AI ethics”. That could include company chats dedicated to AI ethics-related concerns, such as social impact or AI fairness, ethics advisory boards, working groups, philanthropy, data security protocols, codes of conduct, participation processes for employees, training, mission statements, environmental sustainability campaigns, and much more.\footnote{While ethics materials clearly emerged in the data analysis, it has to be noted that we also specifically asked research participants about the processes, roles, activities, which they would classify as “ethics”. We used a very specific thought experiment for that, which we called the “post-it question”, whereby we asked the research participants to mentally visualize their organization of interconnected spheres (spheres could, for example, bind people and processes together by a shared responsibility, such as HR, by responsibility, or by hierarchy, background, interest, or by a product). We then asked them to imagine they had an unlimited stack of post-its with the word “ethics” at hand, and to put an ethics post-it anywhere in the ecosystem of spheres where “ethics” plays a role.} Moss and Metcalf \cite{moss_ethics_2020} term these “ethics methods” and describe them as replicable routines that foster predictability and accountability. 

We, however, want to remain committed to the aforementioned social practice theory approach \cite{shove_dynamics_2012} as an analytical lens which describes “materials” as one of three key elements of any social practice. This lens allows us to examine materials, processes, and technologies that go beyond Moss and Metcalf’s focus on predictability and accountability, and that generally play an important role in the social constitution of AI ethics.

\subsubsection{ Cultural Genealogy} 
In the data analysis, it became clear that the principles we observed were culturally specific and rooted in distinct histories and genealogies of practice. To understand how contemporary AI ethics stabilizes as a social practice and understand its cultural specificity in the German context, it is important to trace the trajectory of the principles, and their interconnectedness with the other elements (especially needs and narratives). We propose doing so by focusing on the cultural genealogy of principles. This is not a matter of mapping traditions to argue for a structuralist notion of cultural reproduction and path dependency, but of mapping the continuity of a social practice and its elements to understand how standardized concepts, such as concepts of “ethics”, emerge in particular contexts.

The cultural genealogies that we trace in our data are not homogeneous, and they do not signify distinct events in Germany’s history. Rather, they trace the social history of concrete materializations of the principles in the context of corporate organizations. These materializations are multidimensional. For example, the principle of “Mitbestimmung” (co-determination) is linked to the cultural history of the “Betriebsrat”, the workers’ council which historically has had significant influence in decision making in German corporate organizations, but also to cultural histories and ideologies of co-determination in general that can be traced back to the Medieval guild system \cite{von_heusinger_von_2010}, and that plays a role in how the collective bargaining power of workers is leveraged via strong unions on the federal level in contemporary Germany. The important part is that even though tech workers in German AI start-ups are not unionized, the thread of the culturally specific continuity of “Mitbestimmung” shapes the professional practice - not as industrial action, but as “ethics”.

\subsection{The Soft Tissue}

\subsubsection{AI Ethics as “Mitbestimmung”}
\hfill\break
{\bfseries AI ethics as “Mitbestimmung” (co-determination) meant employee}\footnote{For this paper, we differentiate between “employee” and “worker” based on the notion that “employees” are permanently employed by an organization, whereby “workers” are precarious \cite{bodie_participation_2014}. The vast majority of individuals we interviewed reported to be permanently employed by their organizations. However, it must be noted that these semantic distinctions are not the same in German, where the terms “Mitarbeiter” or “Arbeitnehmer” designates both “worker”, “employee” or “staff”, and where labor law is much stricter, generally limiting outsourcing and external contracting \cite{haipeter_angestellte_2017}\textbf{.t} participation in corporate decision making.} 

In one of our first conversations, we spoke to a founder of an AI start-up that used computer vision technology to optimize \nobreak production in industrial mechanical manufacturing. Very early in the conversation, it became clear that AI ethics as “Mitbestimmung” drove key decision-making in the organization. Employees had voiced a desire to participate in decisions about what industries and clients the company should and should not work with. Specifically, there was a clear request to be able to refuse work with clients from the military-industrial complex.\footnote{This stance reverberates across the field of German AI start-ups. When talking about ethics, almost every research participant framed refusal to work with the military industrial complex as \textit{ethics}.} Based on that, the company leadership initiated an “ethics council” (consisting of three individuals: an employee, a doctoral candidate [who was not on the payroll] and a student) which conducted a survey among employees to gauge what was of “ethical relevance” to them. This data was then clustered into an “ethics matrix”, a spreadsheet used to vote on taking on new clients or projects. Employees would add their “ethics scoring” regarding a new client or project into the spreadsheet, which were averaged against predetermined thresholds. Projects and clients that scored below the threshold were deemed “okay”, projects just above were deemed to be “in the middle” and needed to be discussed with the ethics council to reach an individual decision, and “extreme cases” were desk rejected by senior management. 

This method was, in part, born out of the need to more deeply explore ethical issues in complex circumstances, particularly in the context of interlinked supply chains for military applications. The overarching goal of how “Mitbestimmung” was operationalized \textit{as ethics} within this organization was to ensure that employees felt that their opinion mattered and was respected, that there was active participation in company decision making (even though the founders still effectively made each decision), and to create transparency across the organization. But it was also conceptualized as a talent attraction and retention tool, and as non-monetary compensation: “We depend on young people joining our company who are willing to work for lower salaries (...) so we need to offer them an environment in which they like to work, because we cannot pay them that much money. (...) But what does environment mean? Where they like to work, where they get a free beer after work and cereal, and also participation (...) real participation.”\footnote{Extended original quote in German: "Und wir sind angewiesen, dass wir junge, talentierte Leute bekommen, die bereit sind für wenig Geld bei uns zu arbeiten. Und dann haben wir so gedacht okay, wie können wir das kompensieren? Also ganz betriebswirtschaftlich. Wir müssen denen halt ein Umfeld bieten, in dem sie gerne arbeiten wollen, weil wir könnten denen nicht soviel Geld bezahlen. Das heißt dann okay, was heißt denn Umfeld? Wo sie gern arbeiten, wo sie ein Feierabendbier kriegen und Müsli... und auch Teilhabe. Und das war halt wirklich Teilhabe, das mit den Kunden, also die moralisch fragwürdigen Deals - findet niemand in Ordnung."}

The {\bfseries needs} that “Mitbestimmung” served were threefold: they derived from the necessity of continually establishing a shared set of values and a shared moral compass. This was connected to the need of establishing transparency about underlying rationales for decision making across the company. A very distinct need that was served through “Mitbestimmung” was the need for talent attraction, alternative compensation, and retention.

The {\bfseries narratives} that underpinned “Mitbestimmung” were:

\begin{itemize}
\item “Our employees need to have a voice in some key company decisions (e.g. client acquisition) to collectively define a shared set of values (‘moral compass’) that will keep the company ethical at large.”
\item “In order to attract and retain the young tech talent that we can ‘afford’ as a start-up, we have to offer ways to be actively involved in ‘ethical’ decision making in the company, because this is a demand in this demographic.”
\item “To keep up morale in our team(s), we need to create transparency about (ethical) decision-making in the company.” 

\end{itemize}

The {\bfseries ethics materials} that substantiated “Mitbestimmung” were:
\begin{itemize}
\item Formalized co-determination processes focused on ethics, such as a platform for employee voting on client selection
\item Routines for “taking temperature” from employees in all-hands team meetings to understand how they “feel” about the direction of the company, specifically with regards to ethical concerns around new projects and clients
\item Regular social gatherings (“Stammtisch”)\footnote{The “Stammtisch” translates to “tribal table” refers to a tradition from Central Europe of regular meetings of a peer group at the same tavern. Historically, these gatherings formed an important public sphere for both for the working class as well as for political activists to share and discuss current affairs in a convivial setting \cite{boyer_spirit_2005}.} to discuss company matters, particularly as they pertain to “ethics” issues, such as social or environmental concerns
\end{itemize}

There is a strong {\bfseries cultural genealogy} of “Mitbestimmung” in the German workplace. It is a well-established corporate governance mechanism in German organizations with a long-standing tradition \cite{spiro_politics_1958}. It has been formalized in a range of laws: initially the “Betriebsrätegesetz” (Worker Council Law) from 1920 to 1933 and now the “Betriebsverfassungsgesetz” (Work Constitution Act) which was introduced in 1952 \cite{muller-jentsch_organisationssoziologie_2003}. Additionally, the “Mitbestimmungsgesetz” (Codetermination Act) from 1976 applies to corporations of 2000 and above employees which has been particularly relevant for the manufacturing sector. This law mandates the representation and participation of employees in the supervisory board (“Aufsichtsrat”) \cite{muller-jentsch_organisationssoziologie_2003}. 

The cultural practices around co-determination in the workplace date back to early days of the industrial revolution in the mid-1800s when, for the first time, large numbers of workers worked for one organization and questions around governance and representation arose. Collective action, famously through strikes of Silesian weavers in 1844 and across the nation in 1848 as well as large strikes among textile and garment workers and miners around the turn of the century \cite{kristof_blacksmiths_1993}, eventually prompted governmental intervention in the young Weimar Republic to formalize the representation of worker interests \cite{abelshauser_vom_1999}. Agreements between company owners, government, and workers materialized in the Weimar “Betriebsdemokratie” (business democracy): in-lieu with the progression of democratic governance overtaking the political arena, “Betriebsdemokratie” was the extension of democratic principles and processes into the realm of the companies permitting the representation of worker interests in a dual manner in society \cite{neumann_freiheit_2015, markovits_politics_2016}. The legacy of these institutions engrained co-determination as guiding governance principle into the modern German economy \cite{plumpe_betriebliche_1999}, for example through the implementation of workers’ councils ("Betriebsrat" or “Betriebsräte”),\footnote{”Betriebsräte” are worker councils that are a formalized representation of workers’ interests within a company. This working relation is mandated by the “Betriebsverfassungsgesetz” (Works Constitution Act)\cite{federal_ministry_of_labour_and_social_affairs_germany_bmas_1972}. A representative or multiple representatives are elected by all employees usually for a period of four years. The “Betriebsrat” is a voluntary position. Employees pursue those commitments on company time and receive full pay. They are tasked with representing employee interests at the top level of the company, spanning from arrangement of health and safety at the workplace, matters of company orders (“Betriebsordnung”, see below), organization of working hours and breaks to the introduction and application of technologies that surveil and evaluate the behavior and performance of employees).} which to this day play a prominent role in German corporate organizations \cite{schnabel_betriebliche_2020}.\footnote{Today ,”Betriebsräte” are in decline. Empirical research suggests that “Betriebsräte” have lost their appeal for  employees as small company-sizes permit a direct negotiation with their management and that they are less likely to be established in owner-managed companies \cite{schnabel_betriebliche_2020}, which often is the case in start-ups. 
} Co-determination came under attack during the Nazi era, when these institutions were targeted, broken-up and replaced with authoritarian governance structures to stifle resistance and opposition \cite{milert_zerschlagung_2014}. Consequently, re-instituting co-determination through fostering strong unions and worker councils was a central mechanism in the rebuilding of postwar Germany \cite{markovits_politics_2016}. 

Today, practices of co-determination are also prominent in the German “Mittelstand”\footnote{The “Mittelstand”, small and medium-sized companies that form an economic unit, is a phenomenon specific to German-speaking countries. The “Mittelstand” is often characterized by family ownership and leadership \cite{berghoff_end_2006}. Whilst business owners hold control over company decisions over its trajectory, they equally carry the responsibilities that come with entrepreneurial ventures \cite{gantzel_wesen_1962} engendering a pronounced emotional attachment and personal commitment to the company \cite{pahnke_german_2019}. This conservative mindset also materializes in strong social bonds between owners and employees and company strategy which are guided by paternalism and longing for multi-generational continuity \cite{berghoff_end_2006}. Traditionally, internal company processes and affairs are characterized by a “patriarchal culture and informality” that promote flat hierarchies between management and staff and establish cooperation via trust and a ‘give and take’-relation \cite{berghoff_end_2006}. These social ties also engrain a strong sense of responsibility for staff which guides a desire for stability and a preference to steady company growth due to maintaining financial autonomy and careful company strategies \cite{berghoff_end_2006}} culture in which familial-like relations between business owners and employees inform a working environment of cooperation, trust, and worker participation \cite{berghoff_end_2006}. While worker organizing, representation, and co-determination is strong in many industries across Germany, and specifically those in the manufacturing sector, employees in AI or tech start-ups often face precarity and tend to be not unionized \cite{dgb_pionierarbeit_2020, beitzer_new_2020}. This is a trend that is different in other parts of the world, for example the U.S. which has seen a growing “tech worker movement” (see, for example \cite{tarnoff_making_2020}), or the Israeli tech industry, which has seen growing unionization \cite{dirksen_trade_2021}.

\subsubsection{AI Ethics as “Verantwortung”}
\hfill\break
{\bfseries AI ethics as both individual responsibility and standardized processes to ensure ethical conduct, as well as corporate responsibility.}

Across almost all of our conversations, the topic of individual and collective responsibility played a central role in how participants constructed and enacted “AI ethics”. In the case of individual “Verantwortung”, one data scientist employed at an AI start-up explained that they trusted their team members to flag ethical concerns if they “sensed” issues. This approach was echoed by another research participant who described regular feedback meetings that served as the space where employees were expected to raise (ethical) issues and concerns they identified within the company.

Collective “Verantwortung” in AI start-ups very concretely materialized in working groups. Generally, these working groups emerged “bottom up” with teams organizing informal gatherings to discuss ethical topics in relation to their organization or their professional practice, ranging from internal company processes, work replacement via automation, to sustainability. For some start-ups, this was time employees spent working on these issues on a voluntary basis, for example during lunch, or after work in a casual setting, such as the company kitchen. Here, employees would be cooking together and discussing ideas pertaining to product development, but also very concrete ethical concerns, such as risk of injury with sensor detection for automated doors, which were then shared as talking and action points with the rest of the company via an internal newsletter. Other start-ups put their employees on the clock to work on ethical issues on company time. For example, one start-up employee explained to us that they were part of a voluntary working group that was tasked with developing a comprehensive ethics strategy for the growing company. For this work, employees were allowed to spend company hours on the work, which was organized in a formal way: some employees would conduct research, others draft parts of the strategy, and findings were presented to the whole company (including senior management) in regular meetings. This research participant remarked that those employees who were part of this working group were representative of those populations most affected by potential AI harms, such as women or queer people. They stated that “it is precisely these groups of people [...] [who join the group] where I think a lot of problems exist [...], where a lot of effort needs to go, because they are not in the majority and questions appear.”\footnote{“[...] das sind halt genau diese Gruppierungen [...], wo ich halt finde, wo es noch ganz viele Probleme gibt. [...] wo halt einfach Arbeit geleistet werden muss, weil das nicht die Mehrheit ist und dort halt eben dann Fragen aufkommen.”}

The {\bfseries needs} that “Verantwortung” satisfied were different from the needs that were satisfied by “Mitbestimmung”. They responded to and evolved around ensuring cohesion across the workforce, and across hierarchies. It was assumed social bonds would be strengthened through the notion of individual and collective “Verantwortung” (explicitly including issues pertaining to the environment), and that this would also serve as conduit for more responsible conduct across and beyond the organization. Relatedly, “Verantwortung” also derived from the need of establishing and maintaining “order” by way of implicit rules, and a culture of compliance, especially vis-à-vis existing European and German data protection regulation. At the same time, “Verantwortung” was seen as a marketing device, serving the need to signal reliability and responsibility to existing and potential clients, which was felt acutely in the context of AI as an emerging and largely poorly understood technology.

The {\bfseries narratives} that underpinned “Verantwortung” were:

\begin{itemize}
\item “We all have an individual responsibility to raise any ethical concerns we may have.”
\item “The individual responsibility to flag ethical concerns serves as necessary checks and balances for corporate conduct and decision making, and as ‘ethics insurance’.”  
\item “Responsible AI means grounding AI development and application in research, and to be transparent about the technology that is being designed.”
\item “Signaling responsibility is key for success-oriented interaction with (potential) investors, clients, and employees, as well as regulators.” 
\item “To ensure individual responsibility and ethical conduct, we formalize our responsibilities, tasks, and commitments with clear frameworks into an organizational order that is transparent to all.” 
\item “Ethics means to fully comply with relevant regulation through standardized processes and responsibilities, particularly in the context of EU and German data protection regulation.” 
\end{itemize}

The {\bfseries ethics materials}  that substantiated “Verantwortung” were: 

\begin{itemize}
\item Internal and external guidelines, such as mission statements, outlining values that ought to inform and steer responsible behavior
\item Internal and external (ethics) advisory boards  
\item Internal chats focused on “ethics” and societal issues, as well as employee time set aside for strike action (specifically the weekly “Fridays for Future” climate strikes)\footnote{“Fridays for Future” is an international youth-led climate-strike movement that started in 2018. Strikes take place during compulsory attendance at school \cite{noauthor_fridays_nodate}.} or collaborative and employee-driven working groups and initiatives to work on and present ethics focused topics, and/or to define mission, strategy or principles, such as ethics principles pertaining to the organization at large organization or to product development
\item Memberships in ethics-focused working groups of industry associations (such as the German AI Start-Up Association) or other organizations (such as United Nations Global Impact), as well as local philanthropic support  
\item Active engagement in academic research through conference participation or research collaborations with university researchers (incl. joint research grants), working with open source models 
\item Compliance officers and/or clearly designated compliance responsibilities across different domains, particularly data protection regulation, but also sustainability, supply chain, HR, or industry-specific regulation (such as in manufacturing or the medical field) 
\end{itemize}

The notion of individual and collective responsibility in the context of company is deeply rooted in German culture \cite{berthoin_antal_rediscovering_2009, palazzo_us-american_2002}. On the one hand, it is tied to the leadership figure of the business owner, or entrepreneur (“Unternehmer”): to “do good” was, and often still is, a religiously motivated virtue among prominent German business leaders, and it includes treating workers well, and increasingly a concern for environmental protection \cite{berghoff_end_2006, berthoin_antal_rediscovering_2009}. Owners of small and medium sized firms are often locally involved in philanthropic efforts \cite{pahnke_german_2019, lehrer_germanys_2015}. The responsibility of the entrepreneur (“Unternehmerverantwortung”) is directly linked to the principle of the social market economy, installed in postwar Germany, which seeks to tame capitalism and capitalists by way of social policies and regulation to maintain a functioning welfare state \cite{bertelsmann_stiftung_normen_1994, habisch_overcoming_2005}. The responsibility of the worker, on the other hand, is more closely tied to the principle of civil society which centers the idea of a “good political order” in which citizens possess civil liberties that are tied to active (democratic) participation, which is grounded in the willingness to inform oneself politically, to participate in elections, and to take up public offices \cite{deutscher_bundestag_bericht_2002}.\footnote{In democratic society, said behavior cannot be enforced. Therefore, civil engagement in society becomes a “political virtue” which marks a “good citizen” \cite{deutscher_bundestag_bericht_2002} or a “good organization”. Being perceived as “good” in that way was an important market signal to the AI entrepreneurs we spoke to.}  

The idea of ethics as “order” (“Ordnung”) in relation to responsibility also comes to matter in the context of internal protocols, routines, and standardized processes. Often, their purpose is to ensure a shared understanding of what constitutes responsibility and ethical procedures. The cultural genealogy of this “order” can be traced via the “Betriebsordnung”, which today is anchored in German labor law and regulates sociality\footnote{The German word used in this context is “Zusammenleben” which literally means “the living together”, but not just in the sense of cohabitation in a domestic setting, but generally in the sense of being a member of different communities: at home, at work, in the family unit, among friends, and so on. 
} in corporate settings. The “Betriebsordnung” focuses on the “Ordnungsverhalten” (“order conduct”) of employees, which is all conduct that is not directly related to performing a job or related tasks (“Arbeitsverhalten”). Interpretations of this law \cite{betriebsraten_ordnung_nodate} govern, for example, whether or not employees must water plants, or if they are allowed to listen to the radio while at work, but also if they are allowed to accept gifts, drink alcohol, or use company equipment privately. As such, it formalizes what is considered morally acceptable within the social space of an organization and what, therefore, is the individual’s responsibility for compliance. Typically, it also defines levels of fines and punishment in case of non-compliance. 

After having started the systematic dismantling of unions starting 1933 as part of the total centralization of power, the Nazi regime introduced a new law, the “Arbeitsordnungsgesetz” (AOG) which gave employers total power to install a new “Betriebsordnung”. Today, the “Betriebsordnung” cannot solely be determined by the employer, but requires involvement and co-determination via the worker council (“Betriebsrat”) as per the “Mitbestimmungsgesetz” (see above). Start-ups do not necessarily have a “Betriebsordnung” that formalizes what constitutes ethical and responsible behavior, but the cultural legacy of (“Betriebs-") “Ordnung” permeates through the strong notion of “Verantwortung”. For example, even though there is no formalization of employees having to voice their ethical concerns about a product, there is an expectation of it, so much so that it is considered a “checks and balances” system for keeping the whole organization ethical.

\section{DISCUSSION
}
The “anatomy of AI ethics” that we presented is emergent from observations and analyses of the social practice of AI ethics. We now want to propose that creating their own “anatomy of AI ethics” can help actors capture continuities, rather than disruptions and breakages. This can facilitate technical and socio-technical innovation that connects to already existing ways of “doing ethics” and ascribing meaning to them, rather than breaking them. For example, the cultural genealogy element provides a pathway for understanding how “AI ethics” is not only not a new concern, but also that its operationalization is a continuation of already existing practices of “ethics” that are culturally specific.

It is important to note that we are not proposing a normative approach of AI ethics practice, nor a simple checklist. Rather, we are proposing a framework that allows actors their own grounded theory.

Understanding existing strategies for putting ethics into action can facilitate the location of connection points to technical approaches and help actors - AI (start-up) managers, engineers, regulators, researchers, professional associations, and more - decide not only what technical and/or socio-technical innovation they need, but where and how to best integrate it into their organization against the backdrop of rapidly changing regulatory regimes. This approach echoes Rakova et al.’s \cite{rakova_where_2021} call for leveraging existing practices to help actors navigate the duality of algorithmic responsibility and organizational structure. We argue that it can be particularly useful for adapting to risk-based regulatory demands, such as the AI Act proposed by the European Commission \cite{european_commission_proposal_2021}, which follows and tasks “market surveillance authorities”\footnote{A “market surveillance authority” is “an authority designated by a Member State (...) as responsible for carrying out market surveillance in the territory of that Member State” \cite{european_commission_regulation_2019} which means monitoring product conformity and compliance with the existing EU health and safety requirements \cite{european_commission_product_2021}.} with compliance control and enforcement. These authorities must individually assess the risk tier of an AI application, and this assessment forms the basis for imposing different obligations onto AI companies, such as adequate risk assessment, documentation and traceability, human oversight, and more \cite{european_commission_regulatory_2021}. 

To effectively and rapidly comply, and sustainably change the social practice of AI design and deployment vis-à-vis these requirements, we suggest actors use the AI ethics anatomy framework and do so by answering two sets of questions. We outline those below in a research and innovation assessment guide. The first set of questions allows them to map the anatomy of existing AI ethics practice/s. The second set of questions allows them to map and assess existing technical and socio-technical innovation as it is relevant to both the regulatory requirements and their own organizational practices. It prompts them to spell out the type and functionality of any given technical and/or socio-technical innovation (e.g. counterfactual explanations), as well as its aims (e.g. in terms of behavioral change, such as through increasing AI literacy among users who get enrolled into AI systems, or change in policy, such as through improved pathways to recourse). It also prompts them to spell out the connection to the principle/s, needs, narratives, ethics materials, and cultural genealogy previously identified by asking how any given technical and/or socio-technical innovation fits a principle, addresses an existing need, can be embedded into an existing narrative, can materialize as part of existing ethics materials, and connects to the cultural genealogy. To conduct this work, actors can fill-in the following table:
\\
\\

\noindent
\begin{tabular}{|p{2.5cm}|p{5.2cm}|}
\hline
\multicolumn{2}{|c|}{{\bfseries 1. Anatomy of Existing AI Ethics Practice/s}} \\
\hline
Principle/s & What is the existing actionable \hfill \break framework related to AI ethics, and what is its meaning?\\
\hline
Needs & What needs are fulfilled by this \hfill \break framework?
 \\
\hline
Narratives & What narratives underpin the principle/s within a specific organizational context?
\\
\hline
Ethics Materials & 
How does the principle concretely\hfill \break materialize?
 \\
\hline
Cultural Genealogy & What are the culturally specific roots and distinct genealogies of practice that \hfill \break underpin the principle?
  \\
 \hline
 \end{tabular}
   \\
   \\
 \begin{tabular}{|p{2.5cm}|p{5cm}|} 
  \hline
\multicolumn{2}{|c|}{	{\bfseries 	2. Technical and Socio-Technical Innovation}} \\
 \hline
 Type \hfill \nobreak and Functionality & \begin{itemize} 
\item What is the technical / \hfill \break socio-technical innovation?
\item How does it work?
\item	What is the aim of this technical / socio-technical innovation?
	\vspace*{-12pt}\end{itemize} \\
	\hline
	Integration & \begin{itemize} 
\item How does it fit the XXX \hfill \break principle?
\item How does it address the existing need of XXX?
\item How can it be embedded into the existing narrative of XXX?
\item How can it materialize as part of existing ethics materials?
\item How does it connect to the \hfill \break cultural genealogy? 
	\end{itemize} \\
\hline
\end{tabular}
\\

Following the research and innovation assessment guide and answering these two sets of questions will prompt actors to reflect on their own existing practices and more critically examine technical and socio-technical innovations that promise to foster fairness, accountability, and transparency vis-à-vis their potential for adoptability. This is because they will begin to understand technical and socio-technical innovation \textit{as practices}, too, which underscores the flexibility of the technical approaches themselves and their vast potential to flexibly be integrated into existing ecosystems of AI ethics practices to help respond to emerging regulatory obligations.

\section{Conclusion}

In the beginning of this paper, we suggested that there is a larger split in the field of AI ethics with rapid growth in works focused on computational interpretations of ethical concerns, normative frameworks, and socio-technical innovation that is outpacing the production of empirical research on the professional practices of AI design and AI ethics specifically. This dynamic produces an epistemological blind spot of the stabilization of social practice and the cultural context of AI and can perpetuate a top-down mentality of finding ways for addressing and mitigating AI harms - whether through behavioral change, or technological innovation \cite{sloane_inequality_2019} - slimming the chances of adoption.

In this paper, we set out to address this issue. We built on social practice theory and on empirical data from our study on the operationalization of “ethics” in German AI start-ups to address the need for a framework that connects technical and socio-technical innovation to qualitative ways of understanding “ethical AI” practices. We have argued that this can help researchers and practitioners more rapidly adapt to changing regulatory regimes pertaining to AI.

Before we make suggestions on future directions of work, we want to reflect our approach, in addition to some of the limitations that have been outlined in the methods section. In addition to those, we want to offer critical reflection on the conceptual approach proposed here. First, we want to acknowledge that centering social practice over individual behavior is by no means a silver bullet. We are not proposing to replace important approaches that already exist in the field, such as the notion of intersectionality\cite{crenshaw_mapping_1991, hammack_jr_intersectional_2018} or design inequality \cite{sloane_need_2019, costanza-chock_design_2020}, but to complement them with a view for initiating change from the bottom up, including in policy. 

Second, we want to caution against a reading of the social practice theory deployed in the context of AI design, which de-centers the individual, as a de-centering of individual responsibility, as a-political. Specifically, we want to underline that focusing on the stages of stability and flux AI practices find themselves in does not mean to absolve individual and powerful actors from their unethical behavior, which ranges from union busting \cite{streitfeld_how_2021}, to tax evasion \cite{neate_silicon_2021}, to harassment \cite{conger_uber_2019}. Social practice theory does not replace a focus on power and oppression, but asks us to examine the relationship between how both come to structure \textit{what we do and how we do it} on both a micro- and a macro level. Third, we want to acknowledge that our work, and the approach we propose here, is anchored in our own positionality, and in the culturally specific data that we used as a basis for our grounded theory work. The strong presence of “principle/s” that gave meaning to AI ethics practices, for example, could be read as a feature that is specific to German culture. Based on our grounded theory work and our data, we can neither confirm nor deny such an interpretation. We suggest that such questions underscore the significance of noting the cultural genealogy of practices, which serves as a strategy for critical positionality and culturally specific interpretation of data.

Against this backdrop, we recommend that future research focuses on three different aspects. First, we recommend increasing the production of qualitative empirical research on the professional practices of ethics and AI development across different cultural contexts, and specifically non-Western contexts, and domains of application. We anticipate that impactful innovation in AI fairness, accountability and transparency will have to be context-, culture-, and domain-specific. Second, we recommned refining the methodology and further detailing and strengthening the conceptual framework of social practice theory, especially vis-à-vis notions of politicality, power, intersectionality, and justice. And third, based on the two catalogues of questions, we recommend developing and testing a usable framework for practical implementation. We hope that this agenda will help close the gap between technical, normative, and socio-technical innovation on fairness, accountability, and transparency, and qualitative research on AI design practices and the impact of AI on social life.

\begin{acks}
This work was supported by the German Federal Ministry of Education and Research (BMBF): Tübingen AI Center, FKZ: 01IS18039A. We thank the reviewers for their helpful comments.
\end{acks}

%%
%% The next two lines define the bibliography style to be used, and
%% the bibliography file.
\bibliographystyle{ACM-Reference-Format}
\bibliography{sample-base}

\end{document}